\documentclass[twocolumn, switch]{article} 

\usepackage{preprint}

\usepackage{amsmath, amsthm, amssymb, amsfonts}
\usepackage[numbers, square]{natbib}
\usepackage[utf8]{inputenc}
\usepackage[T1]{fontenc}
\usepackage{xcolor}

\usepackage{booktabs}
\usepackage{nicefrac}
\usepackage{microtype}
\usepackage{lineno}
\usepackage{float}

\usepackage{subcaption}
\usepackage{newfloat}
\DeclareFloatingEnvironment[name={Supplementary Figure}]{suppfigure}
\usepackage{sidecap}
\sidecaptionvpos{figure}{c}

\usepackage{titlesec}
\titlespacing\section{0pt}{12pt plus 3pt minus 3pt}{1pt plus 1pt minus 1pt}
\titlespacing\subsection{0pt}{10pt plus 3pt minus 3pt}{1pt plus 1pt minus 1pt}
\titlespacing\subsubsection{0pt}{8pt plus 3pt minus 3pt}{1pt plus 1pt minus 1pt}

\usepackage{tikz}
\usepackage[hidelinks, colorlinks=true, linkcolor=purple, urlcolor=blue, citecolor=blue, anchorcolor=black]{hyperref}
\definecolor{lime}{HTML}{A6CE39}
\DeclareRobustCommand{\orcidicon}{
	\begin{tikzpicture}
	\draw[lime, fill=lime] (0,0) 
	circle [radius=0.16] 
	node[white] {{\fontfamily{qag}\selectfont \tiny ID}};
	\draw[white, fill=white] (-0.0625,0.095) 
	circle [radius=0.007];
	\end{tikzpicture}
	\hspace{-2mm}
}
\foreach \x in {A, ..., Z}{\expandafter\xdef\csname orcid\x\endcsname{\noexpand\href{https://orcid.org/\csname orcidauthor\x\endcsname}
			{\noexpand\orcidicon}}
}

\title{Efficacy of Transformer Networks for Classification of Raw EEG Data}

\usepackage{xwatermark}
\newwatermark[firstpage, color=gray!60, angle=90, scale=0.32, xpos=-4.05in, ypos=0] {\href{https://doi.org/}{\color{gray}{Publication doi}}}
\newwatermark[firstpage, color=gray!60, angle=90, scale=0.32, xpos=3.9in, ypos=0] {\href{https://doi.org/}{\color{gray}{Preprint doi}}}
\newwatermark[firstpage, color=gray!90, angle=0, scale=0.28, xpos=0in, ypos=-5in] {*correspondence: \texttt{g\_siddhad@cs.iitr.ac.in}}

\usepackage{authblk}

\author[1\thanks{\tt{g\_siddhad@cs.iitr.ac.in}}]{Gourav~Siddhad\orcidA{}}
\author[1]{Anmol~Gupta\orcidB{}}
\author[2]{Debi~Prosad~Dogra\orcidC{}}
\author[1]{Partha~Pratim~Roy\orcidD{}}

\affil[1]{Department of Computer Science and Engineering, Indian Institute of Technology, Roorkee, Uttarakhand 247667 India}
\affil[2]{School of Electrical Sciences, Indian Institute of Technology, Bhubaneswar, 752050 India}

\begin{document}

\twocolumn[ 
  \begin{@twocolumnfalse} 

\maketitle

\begin{abstract}
With the unprecedented success of transformer networks in natural language processing (NLP), recently, they have been successfully adapted to areas like computer vision, generative adversarial networks (GAN), and reinforcement learning. Classifying electroencephalogram (EEG) data has been challenging and researchers have been overly dependent on pre-processing and hand-crafted feature extraction. Despite having achieved automated feature extraction in several other domains, deep learning has not yet been accomplished for EEG. In this paper, the efficacy of the transformer network for the classification of raw EEG data (cleaned and pre-processed) is explored. The performance of transformer networks was evaluated on a local (age and gender data) and a public dataset (STEW). First, a classifier using a transformer network is built to classify the age and gender of a person with raw resting-state EEG data. Second, the classifier is tuned for mental workload classification with open access raw multi-tasking mental workload EEG data (STEW). The network achieves an accuracy comparable to state-of-the-art accuracy on both the local (Age and Gender dataset; 94.53\% (gender) and 87.79\% (age)) and the public (STEW dataset; 95.28\% (two workload levels) and 88.72\% (three workload levels)) dataset. The accuracy values have been achieved using raw EEG data without feature extraction. Results indicate that the transformer-based deep learning models can successfully abate the need for heavy feature-extraction of EEG data for successful classification.
\end{abstract}
\keywords{Age and Gender \and Classification \and Deep learning \and EEG \and Mental workload \and Transformer network}
\vspace{0.35cm}

\end{@twocolumnfalse}]



\section*{Impact Statement}
The performance and automated feature extraction capabilities of deep learning have seen an enormous rise in recent times. However, in the domain of electroencephalography (EEG), end-to-end deep learning frameworks capable of automated relevant feature extraction are still missing. In this paper, we adapt the highly effective Transformer networks for working with raw EEG data and propose an end-to-end classification framework without the need for hand-crafted feature extraction. We not only show the efficacy of our framework on a locally collected dataset but also that it performed better than state-of-the-art on a public dataset.


\section{Introduction}
\label{sec:introduction}

The field of natural language processing (NLP) witnessed unprecedented success with the introduction of transformer network by Vaswani et al. \cite{vaswani2017Attention}. Radford et al. \cite{radford2018Improving} soon scaled the original transformer network to generative pre-trained (GPT) language models with billions of parameters. They achieved incredible success across different NLP tasks, including question-answering, translation, and many more. In the latest iteration, GPT-3 \cite{brown2020Language}, the authors scaled the parameters to 175 billion and achieved state-of-the-art results. Recently, Fedus et al. \cite{fedus2021Switch} introduced switch-transformers with a trillion parameters and observed substantial gains over the state-of-the-art in over 100 languages. Transformer network has not only proved to be a resounding success in NLP, but it has also been successfully adapted to the field of reinforcement learning \cite{vinyals2019Grandmaster}, computer vision \cite{dosovitskiy2020Image}, and even in GANs \cite{jiang2021Transgan}.

This paper explored the use of transformer networks to classify raw electroencephalography (EEG) data. EEG has long been used for different purposes like disease diagnosis \cite{seal2021deprnet}, BCI \cite{chen2019combination}, neuro-feedback \cite{dousset2020preventing}, workload estimation \cite{zheng2017multimodal}, motor imagery classification \cite{sadiq2021toward}, and even brain-to-brain communication \cite{perez2017brain}. EEG records the electrical activity at the human scalp produced by the ionic currents resulting from simultaneous activation of multiple neurons in the brain \cite{henry2006Electroencephalography}. There are several advantages of using EEG to measure brain activity, including but not limited to its low cost, portability, and being non-invasive with a very high temporal resolution \cite{hamalainen1993Magnetoencephalography}.

However, EEG has several disadvantages as well. The most important of which are noisy raw data \cite{puce2017Review}, high inter-subject variability \cite{lotte2018Review}, and low spatial resolution \cite{kondylis2014Detection}. Artifacts from many non-physiological sources such as power line, bad electrode contact, broken electrodes, etc. and physiological sources such as cardiac pulse, muscle activity, sweating, movement, etc., contaminate raw EEG data \cite{michel2019EEG}. Due to these limitations, pre-processing raw data in EEG plays a vital role and it usually serves as a bottleneck for any classifier. The input fed to the classifier is usually cleaned, filtered, and pre-processed data. Even then, the classifiers could not extract the relevant features as the pre-processing usually results in loss of a significant amount of information and artifact correction cannot recover the EEG signal from the noise in a satisfactory way \cite{pedroni2019Automagic}.

With the recent advances in deep learning \cite{lecun2015Deep}, one expects that the classifier should learn the required features automatically and not rely on hand-engineered features. However, because of the limitations of EEG, most studies have used hand-crafted features from EEG data, even when classifying with complex neural network architectures like deep recurrent convolutions, deep-stacked autoencoders, frequency-dependent CNN, spatial-frequency CNN, etc. \cite{gao2020Complex}.

To build an end-to-end classification system for EEG, we chose to explore the highly effective transformer networks. To better evaluate the efficacy of transformer networks in classifying the raw EEG data, we used one local (Age and Gender classification) \cite{kaur2019Age} and one public dataset (STEW; workload classification) \cite{lim2018STEW}. 

The contributions of this paper are:
\begin{itemize}
    \item Transformer network is used for raw EEG data (cleaned and pre-processed, no feature extraction) classification. To the best of our knowledge, transformer networks have not been used previously for raw EEG data classification.
    \item The proposed framework achieved a state-of-the-art accuracy of 94.53\% (gender) and 87.79\% (age) on a locally collected dataset of age and gender. It also achieved an accuracy of 95.28\% (two workload levels) and 88.72\% (three workload levels) on a public dataset of workload classification. 
    \item The classification results (accuracy, precision, recall, and F1-score) are also compared with the state-of-the-art attention networks. It has been shown that the transformer networks are a better choice, especially without needing any hand-crafted feature extraction.
\end{itemize}

The rest of the paper is organized as follows. Section \ref{sec:background} presents a review of the latest articles that have classified age, gender, and mental workload with EEG data. Section \ref{sec:methodology} describes the methodology used for the whole experiment, including the task, pre-processing, and the transformer network's details. Section \ref{sec:results} reports the results and the conclusion is given in Section \ref{sec:conclusion}.


\section{Background}
\label{sec:background}

For testing the transformer network, we first selected our local Age and Gender EEG dataset collected by Kaur et al. \cite{kaur2019Age}. The dataset consists of resting-state EEG data of 60 individuals. To further validate the performance of the proposed methodology, it was tuned and applied on the open-access mental workload dataset known as STEW \cite{lim2018STEW}. In STEW, data of 48 subjects were collected while doing a SIMKAP task (details in the Methodology section). We chose STEW dataset since a similar 14-channel EEG headset was used for the collection and the classification paradigm included both binary and multi-class classification identical to the local dataset. In the following subsections, the relevant literature for both datasets is discussed.


\subsection{Age and Gender Classification}

There is limited available literature where EEG data has been used to identify age and gender, especially for age. If the discussion is restricted to deep learning with EEG data, then the field is still very much unexplored. Putten et al. \cite{putten2018Predicting} applied CNN on EEG data to classify gender and achieved an accuracy of 80\%. This might be the most extensive study with EEG data for gender recognition as they collected data of 1308 subjects using 26-channel EEG with 47\% male subjects and 53\% female subjects. Zhang et al. \cite{zhang2019Gender} used the phase locking values (PLV) as a feature for gender classification on the DEAP dataset using extreme learning machines (ELM) and achieved an accuracy of 80.0\%. 

In our previous work, Kaushik et al. \cite{kaushik2019EEG} with the same dataset (Age and Gender), used DWT to decompose the EEG signals into five frequency bands. Deep neural networks, namely long short-term memory (LSTM), bi-directional LSTM (BLSTM), and BLSTM-LSTM, were presented, out of which BLSTM-LSTM gave the best accuracy 97.52\% and 93.70\% on beta frequency band for gender and age classification, respectively. As evident, the field of identifying age and gender using EEG and deep learning is yet to be explored. Kaur et al. \cite{kaur2019Age} pre-processed EEG data using Savitsky-Golay (SG) filter, smoothing is done, and discrete wavelet transform (DWT) decomposition is performed for extracting five frequency bands. Statistical features are extracted from the frequency bands. Then random forest classifier is used, which gave an accuracy of 96.66\% and 88.33\% for gender and age classification, respectively.


\begin{figure*}[!t]
    \centering
    \begin{subfigure}[b]{0.36\textwidth}
        \centering
        \includegraphics[width=\textwidth]{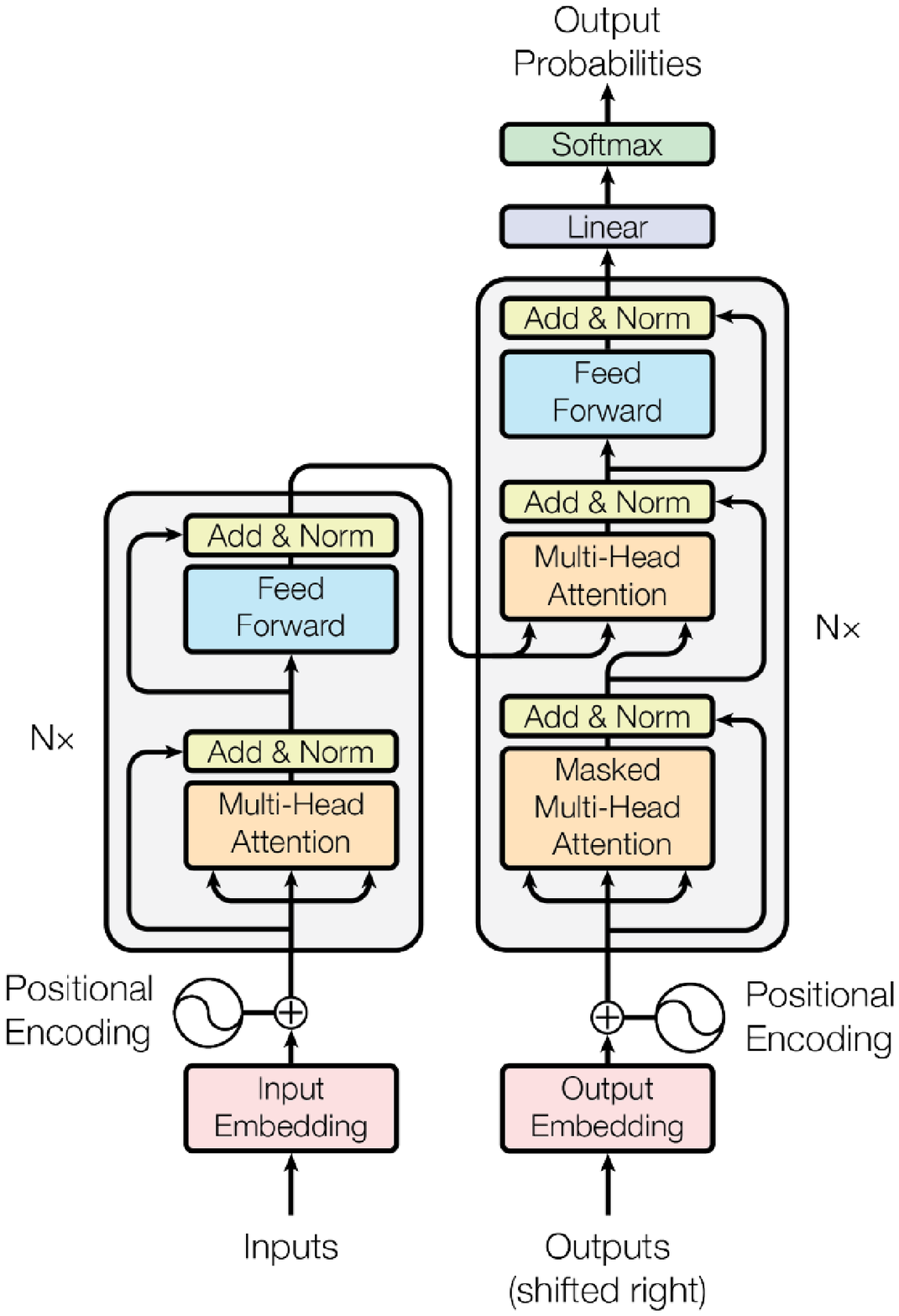}
        \vspace{-0.5cm}
        \caption{}
        \label{fig_arch:tn}
    \end{subfigure}
    \hfill
    \begin{subfigure}[b]{0.14\textwidth}
        \centering
        \includegraphics[width=\textwidth]{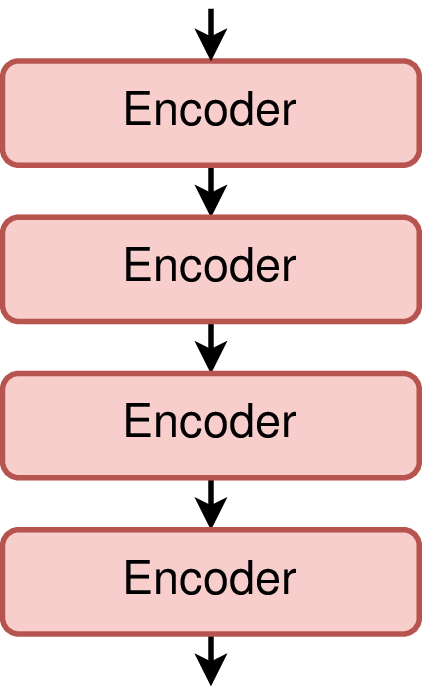}
        \vspace{-0.5cm}
        \caption{}
        \label{fig_arch:enc}
    \end{subfigure}
    \hfill
    \begin{subfigure}[b]{0.16\textwidth}
        \centering
        \includegraphics[width=\textwidth]{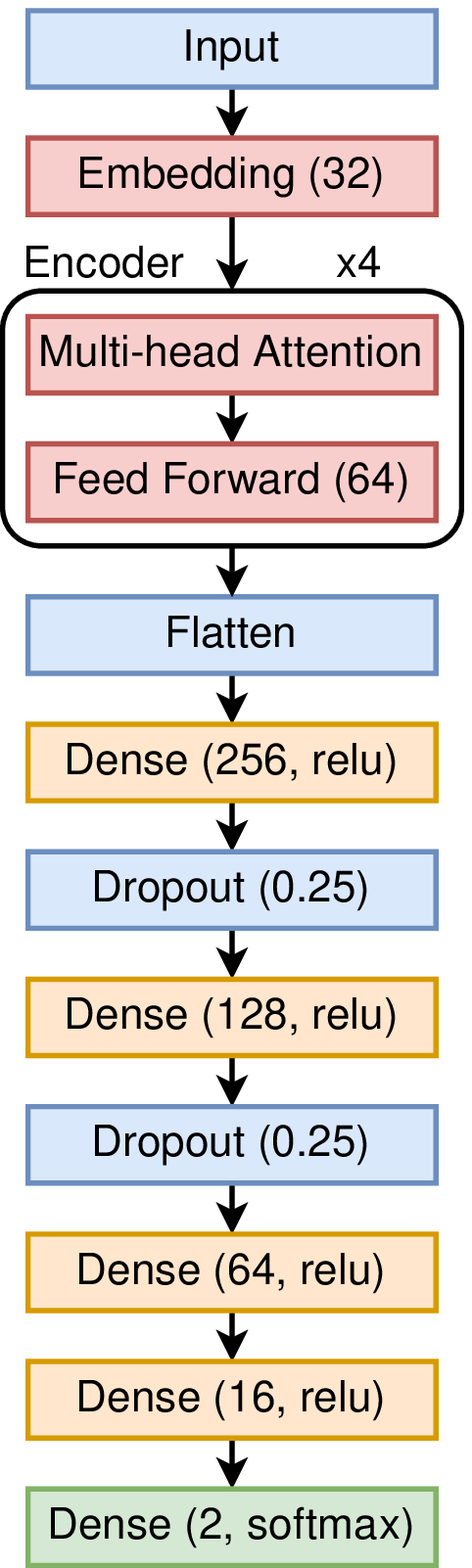}
        \vspace{-0.5cm}
        \caption{}
        \label{fig_arch:ag}
    \end{subfigure}
    \hfill
    \begin{subfigure}[b]{0.16\textwidth}
        \centering
        \includegraphics[width=\textwidth]{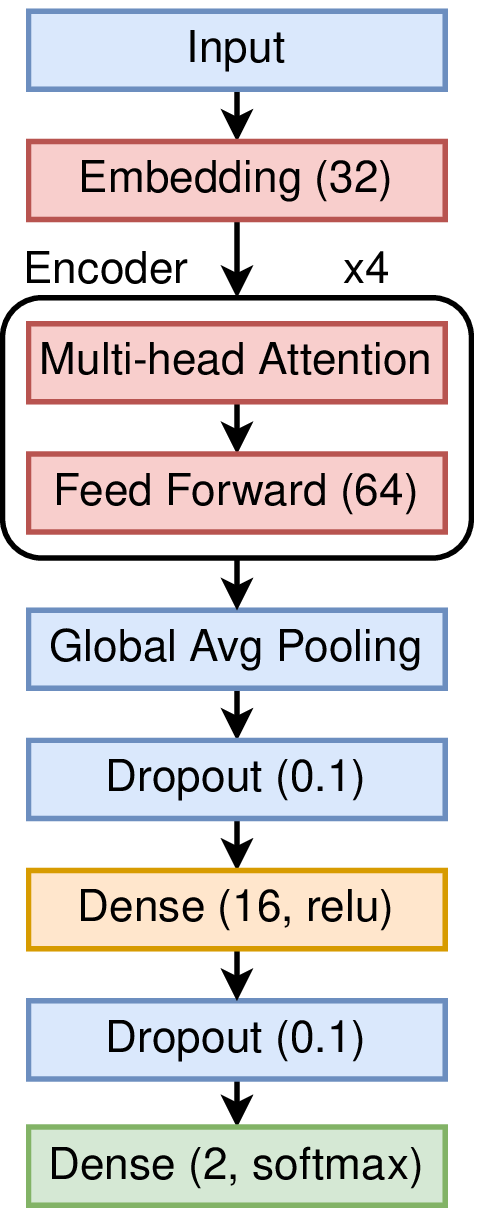}
        \vspace{-0.5cm}
        \caption{}
        \label{fig_arch:is}
    \end{subfigure}
    \caption{Network architecture for (a) Original transformer network \cite{vaswani2017Attention} (left is encoder and decoder is on the right); (b) Arrangement of encoders in transformer block (stack of 4 encoders are used in the proposed method); Proposed architecture for (c) Age and Gender, and (d) STEW dataset.}
    \label{fig_arch}
\end{figure*}

\subsection{Workload Classification}

EEG can help us provide an objective, absolute, and temporally sensitive workload measure. Therefore, it has been used time and again for workload classification. Qu et al. \cite{qu2020Mental} used support vector machine (SVM) for classifying EEG data collected from 13 participants while doing the NASA MATB-II task \cite{espada2011Multiattribute}. They used energy from the independent components extracted from independent component analysis (ICA) as features and achieved an accuracy of 79.8\% for the three workload levels (low, medium, and high). Different machine learning algorithms such as linear discriminant analysis \cite{bagheri2020EEG}, random forest \cite{pei2021EEG}, KNN, and SVM \cite{gupta2020Classification} have been utilised for workload classification from EEG data.

Different neural network architectures have already been shown to classify mental workload successfully. Hefron et al. \cite{hefron2017Deep} used stacked LSTMs for classifying data of six participants using the MATB-II. They achieved an accuracy of 93\%. Sun et al. \cite{sun2021WLnet} proposed WLNet specifically for classifying mental workload and compared it to temporally constrained sparse group spatial patterns (TCSGSP, a deep version of the CSP method proposed by Zhang et al. \cite{zhang2018Temporally}) and EEGNet \cite{lawhern2018EEGNet} (proposed for BCI applications, consisting of depth-wise and separable convolution layers). WLNet was a shallow neural network consisting of one-dimensional convolution layers, thereby requiring fewer parameters. They did binary workload classification on the dataset provided by Fabien Lotte research group at INRIA \cite{lotte2018Review} and found that WLNet out-performed TCSGSP and EEGNet. Saha et al. \cite{saha2018Classification} used a stacked denoising autoencoder to classify three workload levels and achieved 89.51\% accuracy. 

Specifically for the STEW dataset, different neural networks have been applied and compared. Chakladar et al. \cite{chakladar2020EEG} compared the performances of stacked LSTM, BLSTM, CNN-LSTM, stacked autoencoders with LSTM/BLSTM, and their proposed model of hybrid BLSTM-LSTM network. Their proposed model achieved state-of-the-art accuracy of 86.33\% and 82.57\% for binary and tertiary classification on the STEW dataset. Lim et al. \cite{lim2018STEW} made use of artifact subspace reconstruction (ASR) and features were extracted using power spectral density (PSD) and neighborhood component analysis (NCA). Using the support vector regression (SVR) model, the accuracy of 69.20\% was achieved for the ``SIMKAP multi-task'' classification. Zhu et al. \cite{zhu2021Cognitive} constructed oblique visibility graphs (OVG) and frequency domain features (mean degree, clustering coefficient, and degree distribution) were extracted. Decision tree (DT) and SVM were used for classification, where SVM performed better and yielded 89.60\% and 79.50\% for binary and tertiary classification, respectively.

Attention-based networks have also been used for EEG classification. Zhu et al. \cite{zhu2020convolution} proposed a neural network based on CNN and attention mechanism to perform automatic sleep staging. Zheng and Chen \cite{zheng2021attention} proposed an end-to-end attention-based Bi-LSTM method, named Bi-LSTM-AttGW for visual cognition task. The attention weighting method is applied to Bi-LSTM output.

None of the above-cited literature for both datasets worked with raw EEG data. The input fed to the deep neural network has varied from simple band segregated data (alpha, beta, gamma, etc.) to simple features like mean, variance, etc. of the band power to much more complex features like PLV and PLI. Thus, we can conclude that EEG poses a unique challenge of complex pre-processing and feature selection mechanisms.

In this paper, we explored the possibility of using raw EEG data for doing classification. Given the available literature, it was evident that we needed to use a proficient neural network architecture that can automatically extract the relevant features. Recently, transformer networks have seen a phenomenal rise and are used in state-of-the-art accomplishments in almost every domain \cite{dosovitskiy2020Image, jiang2021Transgan}. In the context of EEG, transformer networks have been rarely used but have been shown to generate state-of-the-art results for sleep stage classification \cite{kostas2021bendr} and speech recognition \cite{krishna2020EEG}. Therefore, we decided to use a transformer network for the classification.


\section{Methodology}
\label{sec:methodology}

This section discusses the proposed transformer network for the classification of age and gender and mental workload estimation. After acquisition and pre-processing is done on EEG data after creating epochs. The transformer network is then trained for classification on this raw EEG data.

\begin{figure}[!t]
	\centering
	\includegraphics[width=\columnwidth]{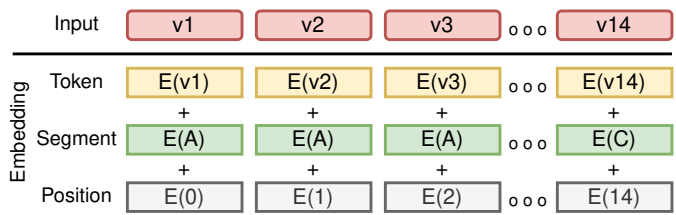}
	\caption{This figure shows the working of embedding and positional encoding. E with numbers represents the position embedding, E with different alphabets represents different segments, and E with various v represents the token vectors for the input. These are sum together to produce the input for the transformer network.}
	\label{fig_epe}
\end{figure}

\begin{figure*}[!t]
    \centering
    \begin{subfigure}[b]{0.24\textwidth}
        \includegraphics[width=\textwidth]{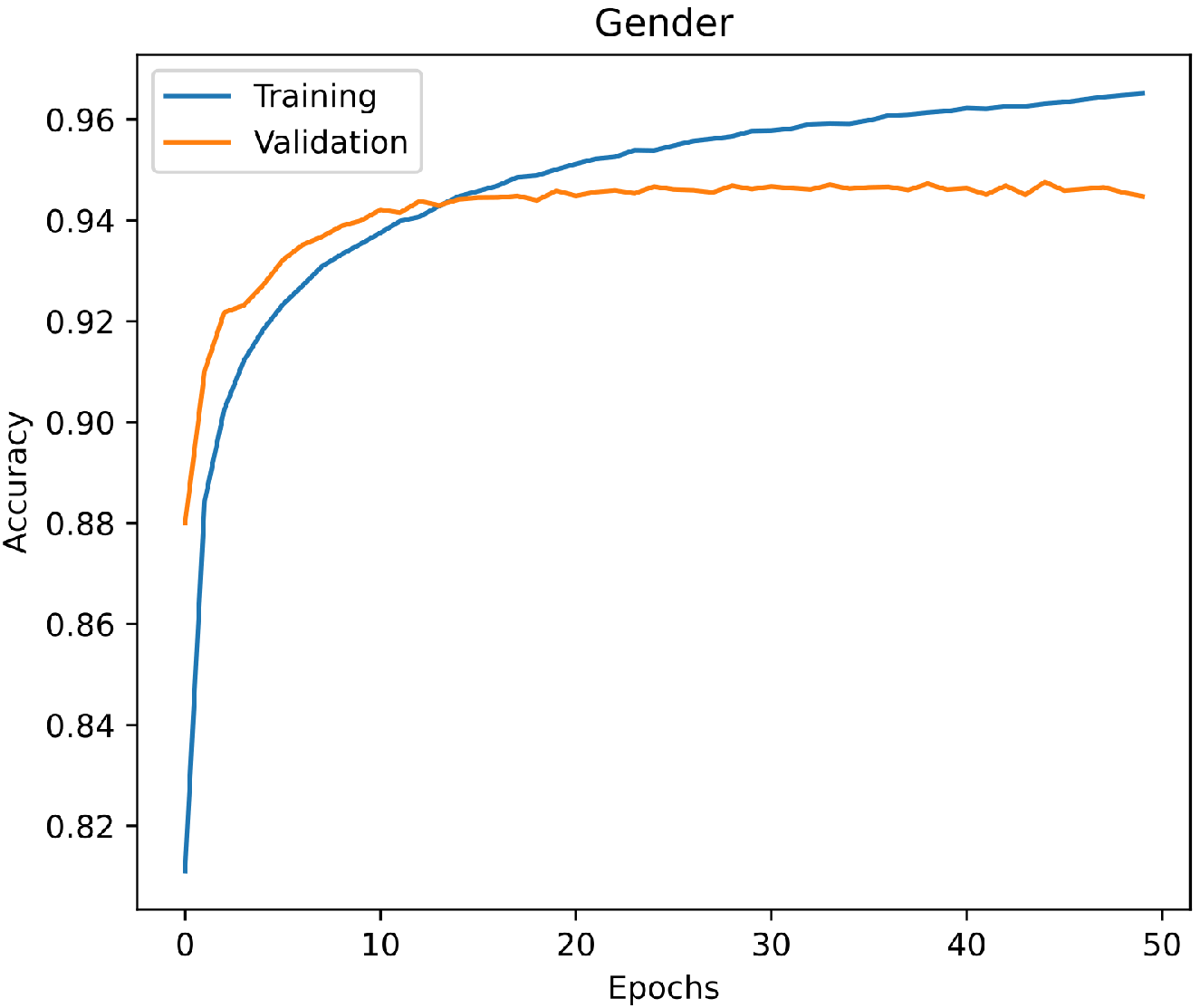}%
        \vspace{-0.2cm}
        \caption{}
        \label{fig_th:ag2c}
    \end{subfigure}
    \begin{subfigure}[b]{0.24\textwidth}
        \includegraphics[width=\textwidth]{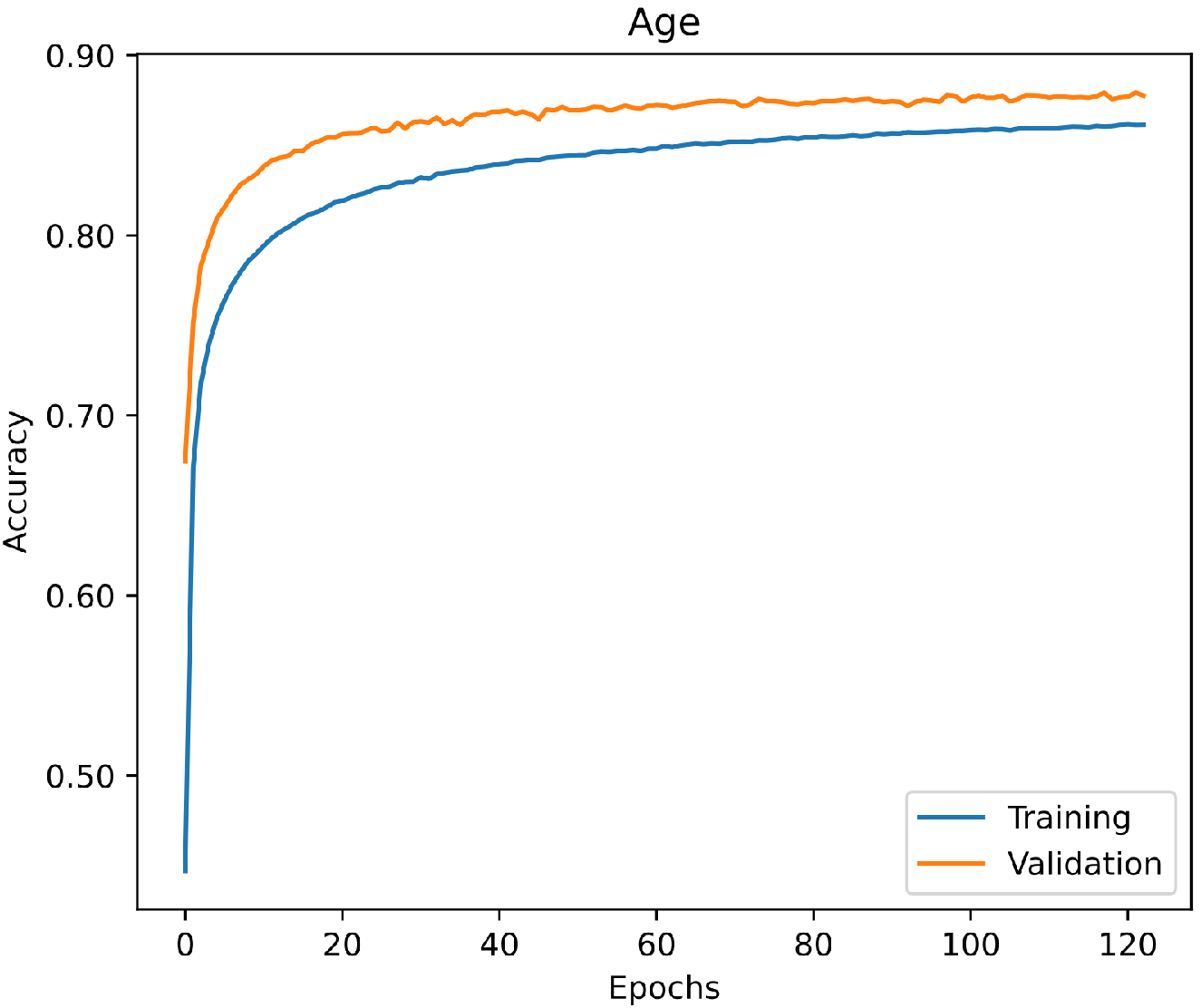}%
        \vspace{-0.2cm}
        \caption{}
        \label{fig_th:ag6c}
    \end{subfigure}
    \begin{subfigure}[b]{0.24\textwidth}
        \includegraphics[width=\textwidth]{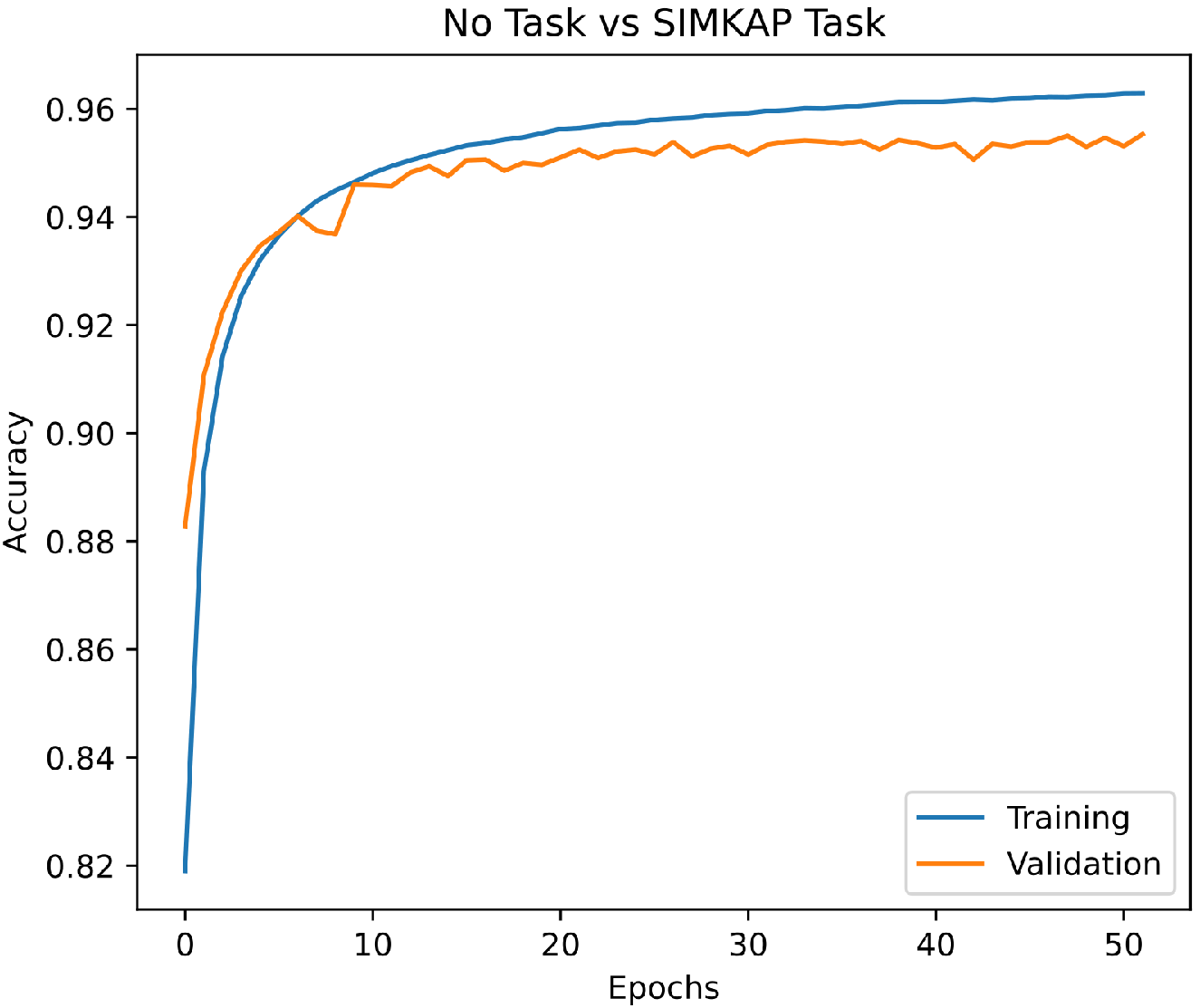}%
        \vspace{-0.2cm}
        \caption{}
        \label{fig_th:is2c}
    \end{subfigure}
    \begin{subfigure}[b]{0.24\textwidth}
        \includegraphics[width=\textwidth]{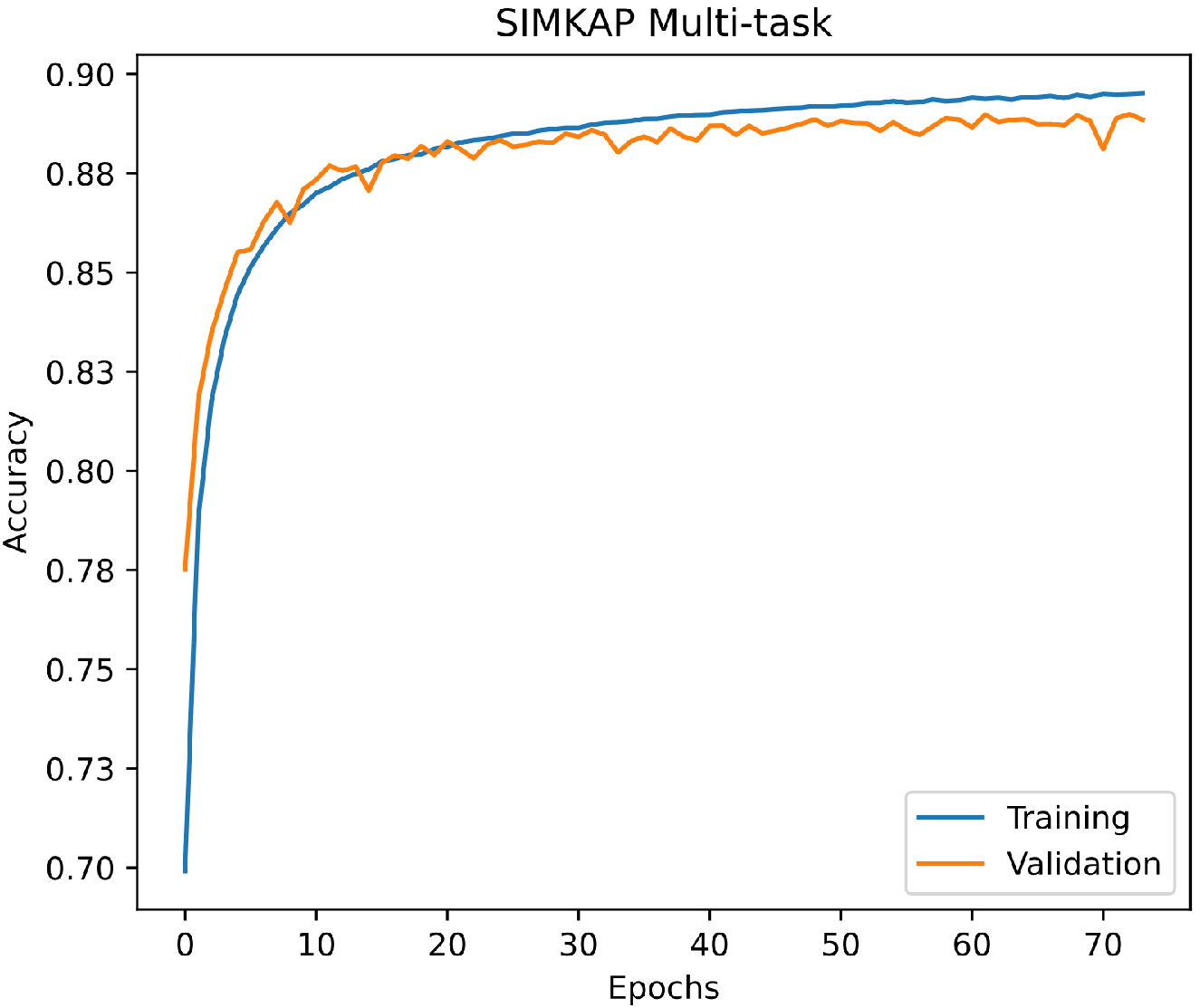}%
        \vspace{-0.2cm}
        \caption{}
        \label{fig_th:is3c}
    \end{subfigure}
    
    \caption{Train and validation accuracy for Age and Gender (a) Gender, (b) Age; and STEW dataset (c) No vs SIMKAP task, (d) SIMKAP multi-task.}
    \label{fig_th}
\end{figure*}

\subsection{Experimental Data}

\subsubsection{Age and Gender Dataset}

This dataset \cite{kaur2019Age} consists of raw EEG signals from 60 users (25 female and 35 male) with an age range between 6 and 55 years. The data was collected with Emotiv Epoch Plus (14 channels) with a sampling rate of 128 Hz. The data was acquired at a resting state for 10 sessions of 10 seconds each.


\subsubsection{STEW Dataset}

The simultaneous task EEG workload (STEW) dataset \cite{lim2018STEW} consists of raw EEG data of 48 subjects who participated in a multitasking workload experiment that utilized the simultaneous capacity (SIMKAP) multitasking test. The signals were captured using the Emotiv EPOC EEG headset, with 16-bit A/D resolution, 128 Hz sampling frequency, and 14 channels, namely AF3, F7, F3, FC5, T7, P7, O1, O2, P8, T8, FC6, F4, F8, AF4 according to the 10-20 international system with two reference channels (CMS, DRL). There are two parts to the experiment. First, the data was acquired for 2.5 minutes with subjects at rest or ``No task''. Next, subjects performed the SIMKAP test with EEG being recorded and the final 2.5 minutes were used as the workload condition. Subjects rated their perceived mental workload after each segment of the experiment on a rating scale of 1-9.


\subsection{EEG and Pre-processing}

EEG signals in their raw form (captured from a device) contain noise and artifacts and need to be cleaned before use. EEG data is imported and bandpass filtering is done to remove environmental/muscle noise in scalp EEG. After epoching and removing bad epochs from the data, independent component analysis (ICA) is applied and bad channels are manually removed. The dataset is used with a sampling rate of 128 Hz, which is the same as during acquisition.


\begin{figure}[!t]
    \centering
    \begin{subfigure}[b]{\columnwidth}
        \includegraphics[width=\columnwidth]{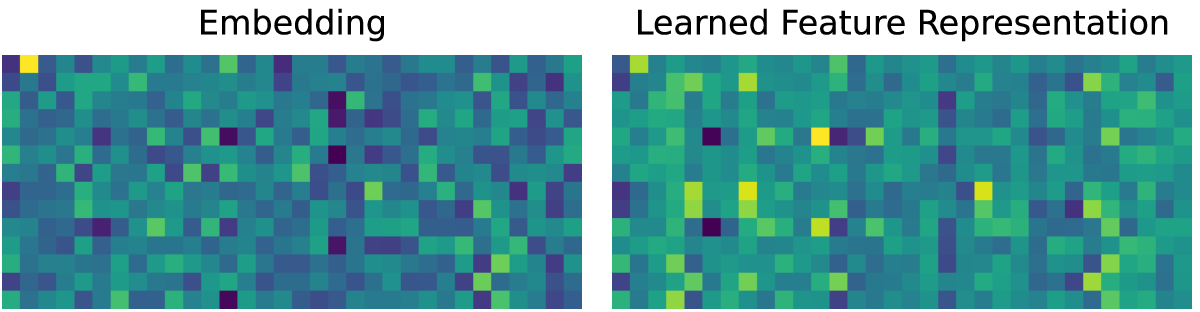}%
        \vspace{-0.2cm}
        \caption{}
        \label{fig_va:ag2c}
    \end{subfigure}
    
    \begin{subfigure}[b]{\columnwidth}
        \includegraphics[width=\columnwidth]{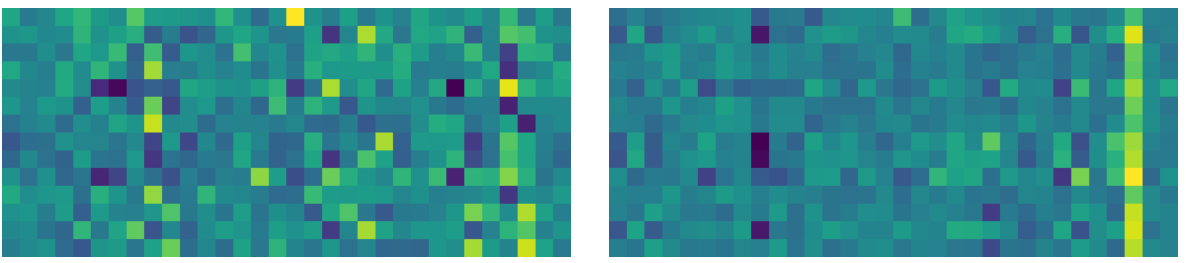}%
        \vspace{-0.2cm}
        \caption{}
        \label{fig_va:ag6c}
    \end{subfigure}
    
    \begin{subfigure}[b]{\columnwidth}
        \includegraphics[width=\columnwidth]{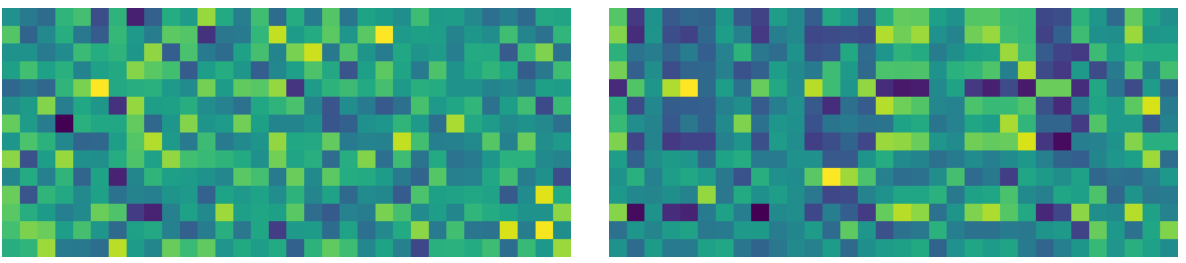}%
        \vspace{-0.2cm}
        \caption{}
        \label{fig_va:is2c}
    \end{subfigure}
    
    \begin{subfigure}[b]{\columnwidth}
        \includegraphics[width=\columnwidth]{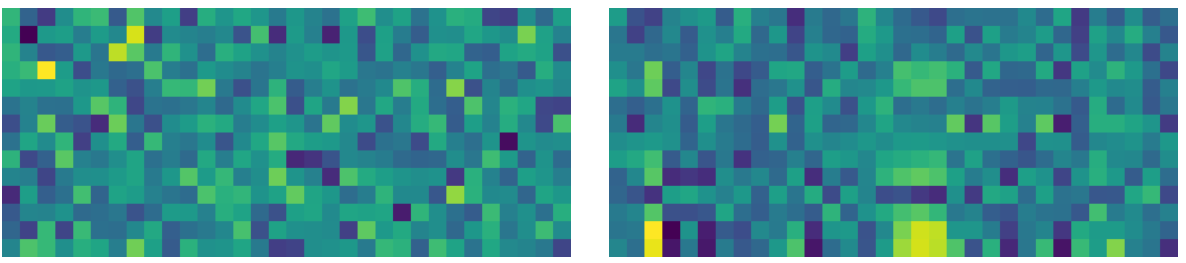}%
        \vspace{-0.2cm}
        \caption{}
        \label{fig_va:is3c}
    \end{subfigure}
    \caption{Visualisation of embedding (left) and learned feature representation (right) for Age and Gender dataset (a) Gender, (b) Age; and STEW dataset (c) No vs SIMKAP task, (d) SIMKAP multi-task.}
    \label{fig_va}
\end{figure}

\subsection{Transformers}

The transformer network \cite{vaswani2017Attention} is a neural network based on a self-attention mechanism. It has proved higher quality and lower computation requirements for language translation tasks than recurrent and convolution models. It applies a self-attention mechanism that directly models relationships between all input parts, regardless of their position. The result of these comparisons is an attention score for every other part in the input, which determines their contributions. The transformer network (shown in Fig \ref{fig_arch:tn}) consists of encoders and decoders. The encoders are stacked in the transformer model as shown in Fig \ref{fig_arch:enc}. Each encoder has two sub-layers, a multi-head self-attention mechanism, and a position-wise fully connected feed-forward network. There exists a residual connection around both sub-layers, followed by a normalization layer. The decoder is not used here.

\begin{table*}[!t]
	\centering
	\caption{Classification analysis for Age and Gender and STEW dataset}
	\renewcommand\arraystretch{1.1}
	\begin{tabular}{l c c c c c c}
		\toprule
        \textbf{Model} & \textbf{Features} & \textbf{Classifier} & \multicolumn{2}{c}{\textbf{Accuracy}} \\

        \midrule
        \multicolumn{3}{l}{Age and Gender Dataset} & Gender (2 class) & Age (6 class) \\
		\midrule

        Kaur et al. \cite{kaur2019Age}
            & Statistical
            & Random Forest
            & 96.66 & 88.33 \\
            
        Kaushik et al. \cite{kaushik2019EEG}
            & Frequency
            & BLSTM-LSTM
            & 97.50 & 93.70 \\
        
        \textbf{Proposed work}
            & \textbf{Not used}
            & \textbf{Transformer}
            & \textbf{94.53} & \textbf{87.79} \\
        
        \midrule
        \midrule
        \multicolumn{3}{l}{STEW Dataset} & No vs SIMKAP task (2 class) & SIMKAP multi-task (3 class) \\
		\midrule

        Lim et al. \cite{lim2018STEW}
            & PSD, NCA 
            & SVR 
            & - & 69.20 \\
            
        Zhu et al. \cite{zhu2021Cognitive}
            & Frequency
            & DT, SVM 
            & 89.60 & 79.50 \\
        
        Chakladar et al. \cite{chakladar2020EEG}
            & Frequency, Time
            & BLSTM-LSTM
            & 86.33 & 82.57\\
        
        \textbf{Proposed work}
            & \textbf{Not used}
            & \textbf{Transformer}
            & \textbf{95.28} & \textbf{88.72} \\
        
		\bottomrule
	\end{tabular}
	\label{tab_results}
\end{table*}
    
\begin{figure*}[!t]
	\centering
	\begin{subfigure}[b]{0.24\textwidth}
	    \centering
		\includegraphics[width=\textwidth]{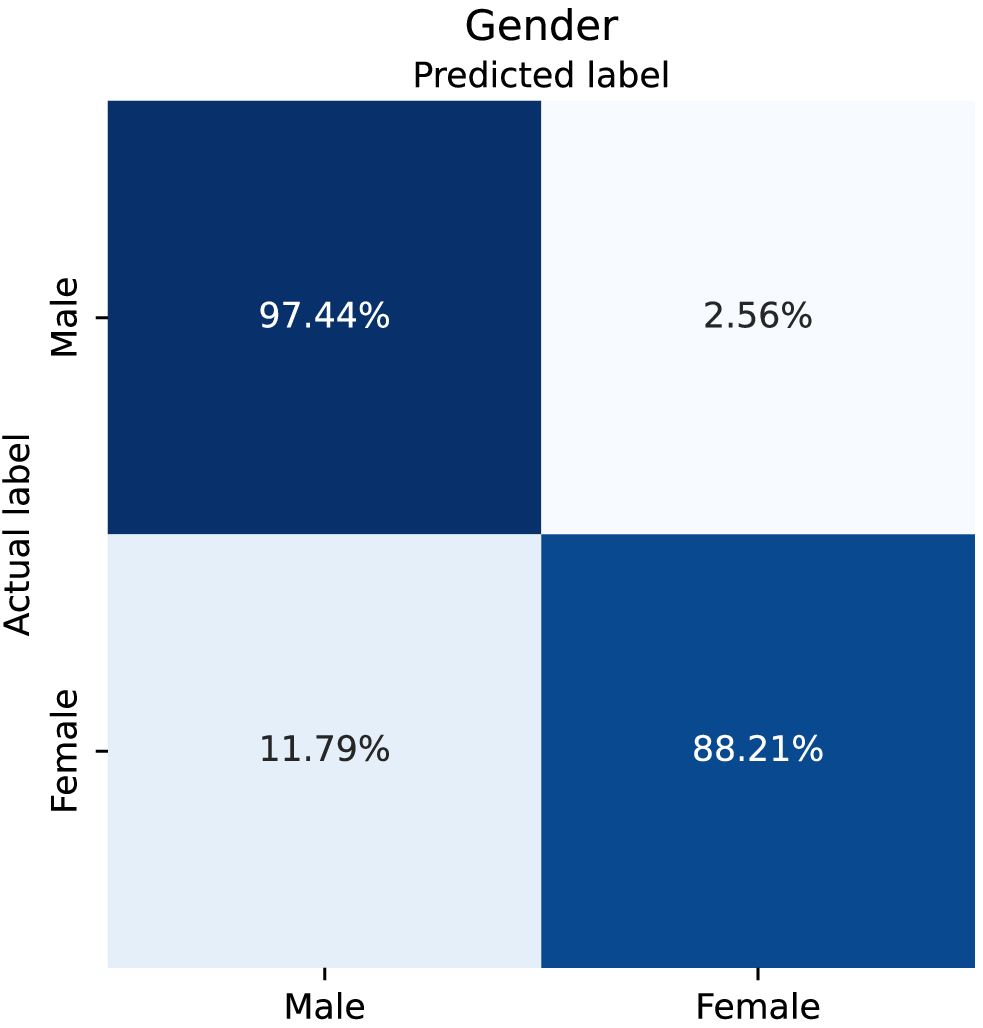}
		\vspace{-0.5cm}
		\caption{}
		\label{fig_cm:ag2c}
	\end{subfigure}
	\begin{subfigure}[b]{0.24\textwidth}
	    \centering
		\includegraphics[width=\textwidth]{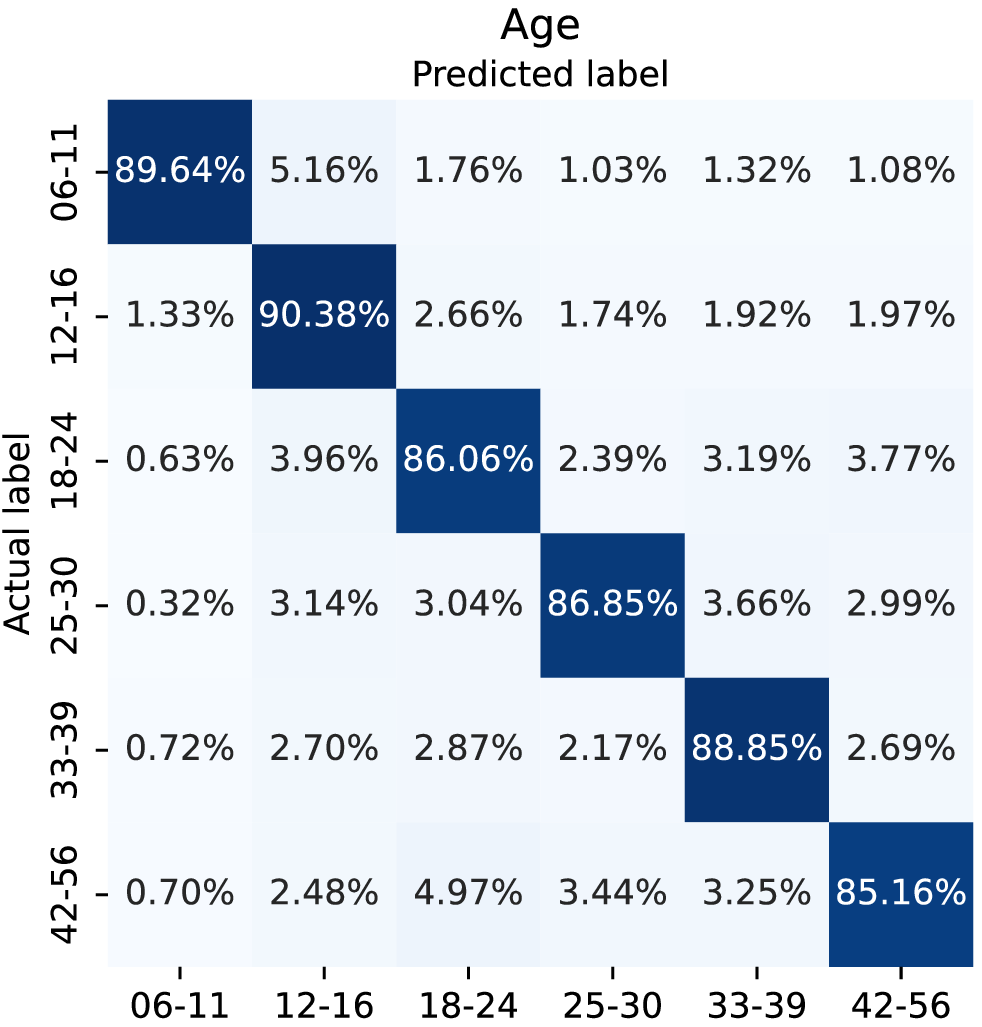}
		\vspace{-0.5cm}
		\caption{}
		\label{fig_cm:ag6c}
	\end{subfigure}
	\begin{subfigure}[b]{0.24\textwidth}
	    \centering
	    \includegraphics[width=\textwidth]{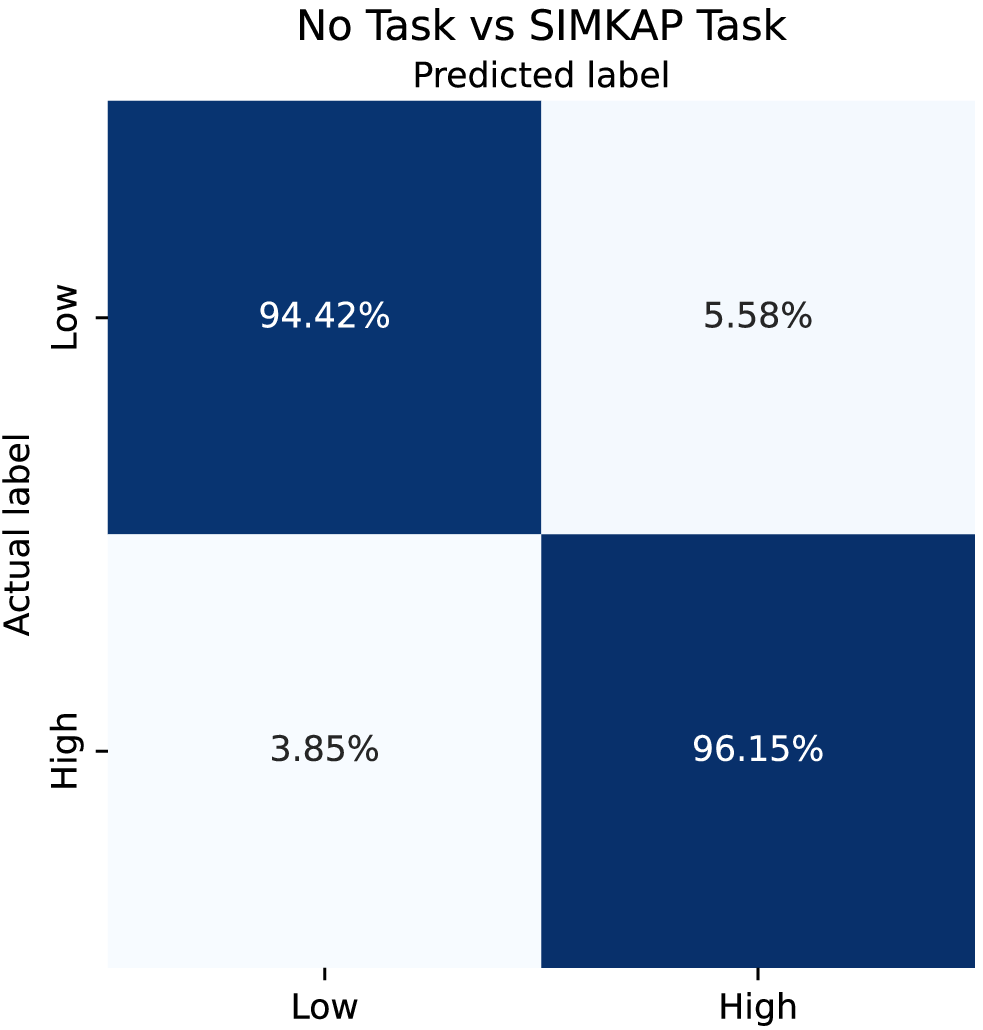}
		\vspace{-0.5cm}
		\caption{}
		\label{fig_cm:is2c}
	\end{subfigure}
	\begin{subfigure}[b]{0.24\textwidth}
	    \centering
		\includegraphics[width=\textwidth]{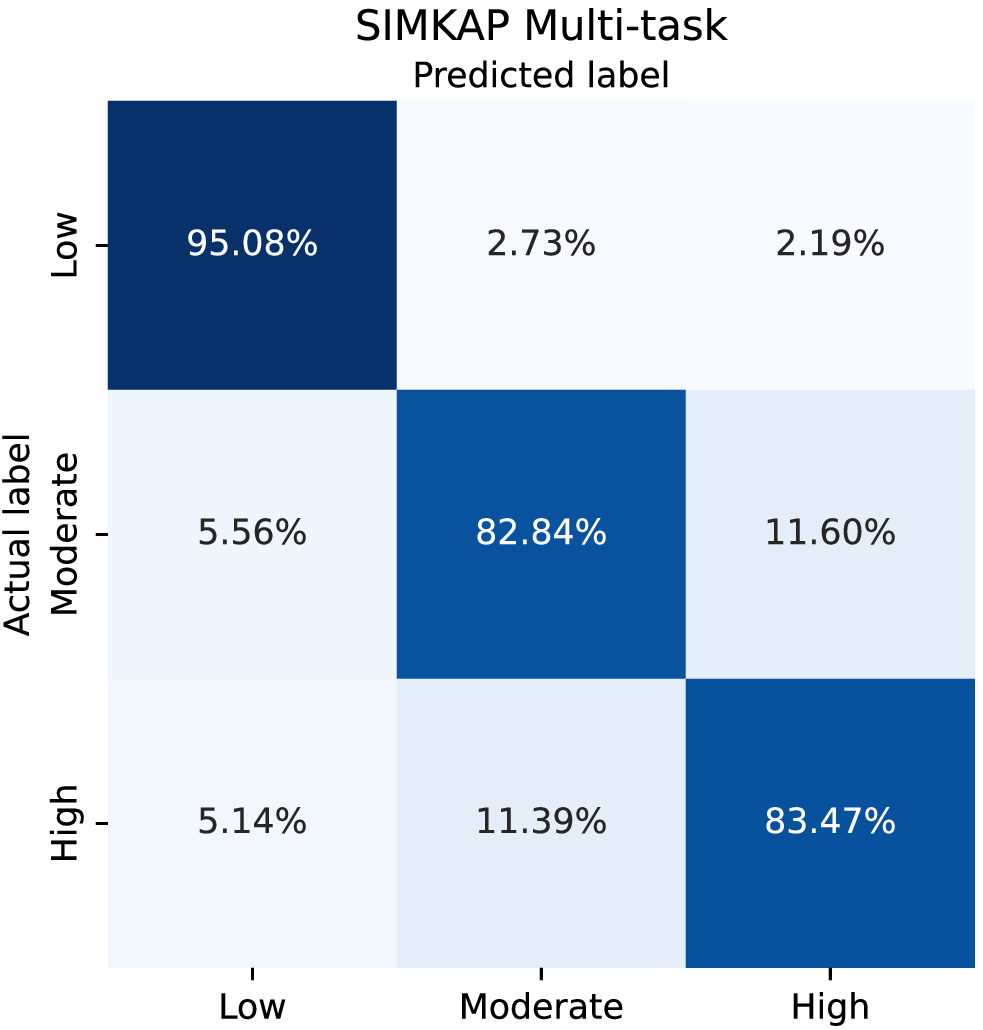}
		\vspace{-0.5cm}
		\caption{}
		\label{fig_cm:is3c}
	\end{subfigure}
	\caption{Confusion matrix for Age and Gender dataset (a) Gender, (b) Age; and STEW dataset (c) No vs SIMKAP task, (d) SIMKAP multi-task.}
	\label{fig_cm}
\end{figure*}

\begin{figure*}[!t]
	\centering
	\begin{subfigure}[b]{0.24\textwidth}
	    \centering
		\includegraphics[width=\textwidth]{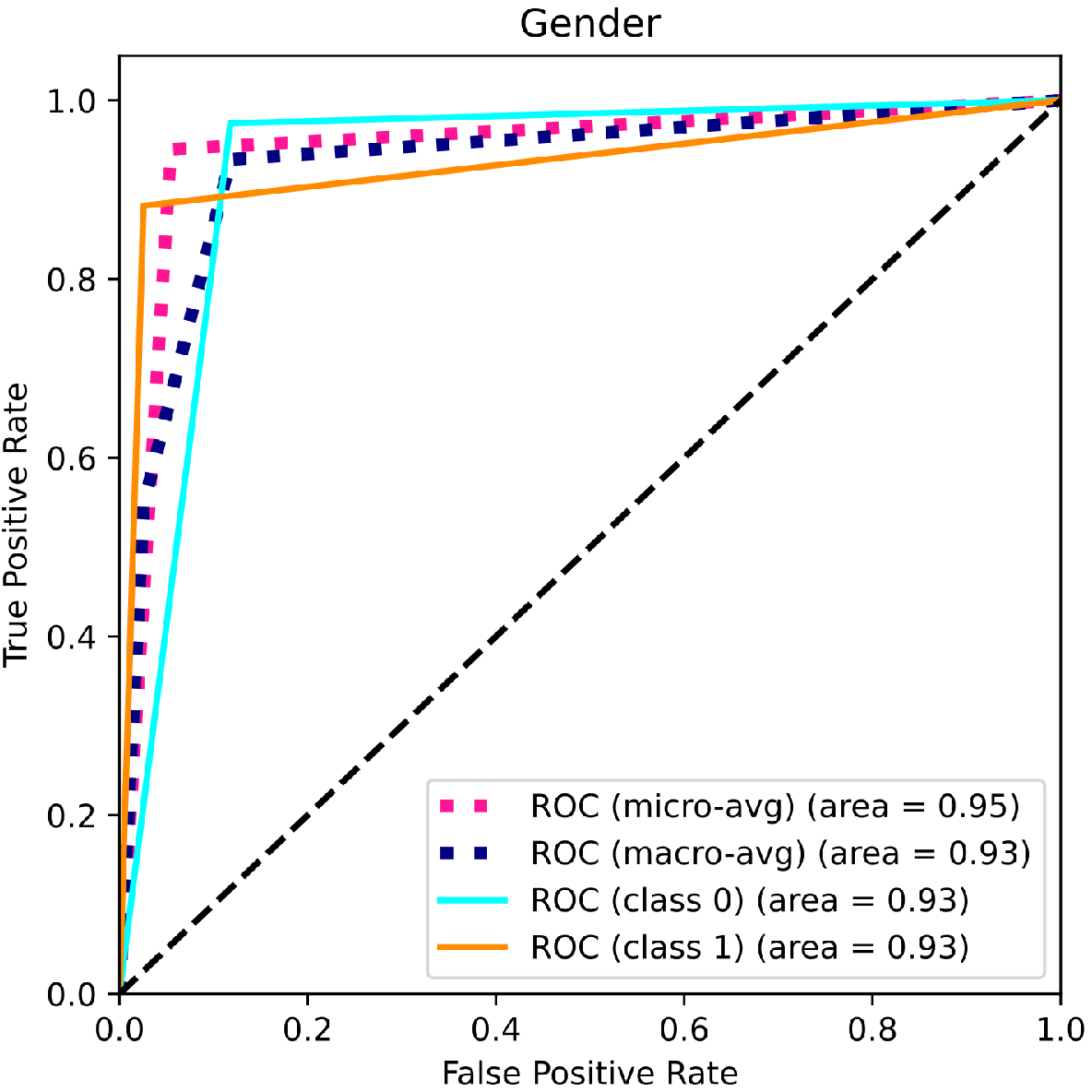}
		\vspace{-0.5cm}
		\caption{}
		\label{fig_roc:ag2c}
	\end{subfigure}
	\begin{subfigure}[b]{0.24\textwidth}
	    \centering
		\includegraphics[width=\textwidth]{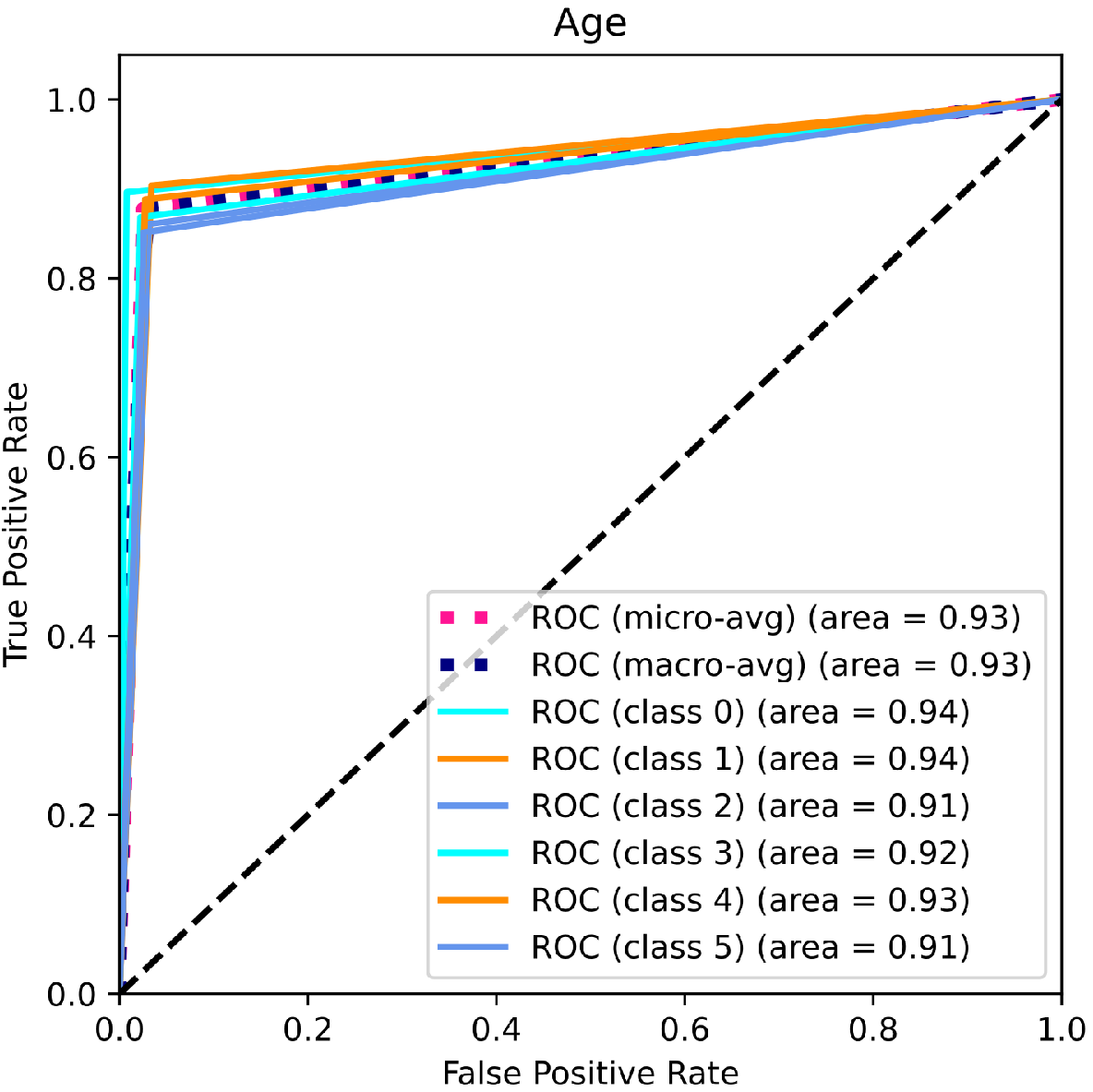}
		\vspace{-0.5cm}
		\caption{}
		\label{fig_roc:ag6c}
	\end{subfigure}
	\begin{subfigure}[b]{0.24\textwidth}
	    \centering
		\includegraphics[width=\textwidth]{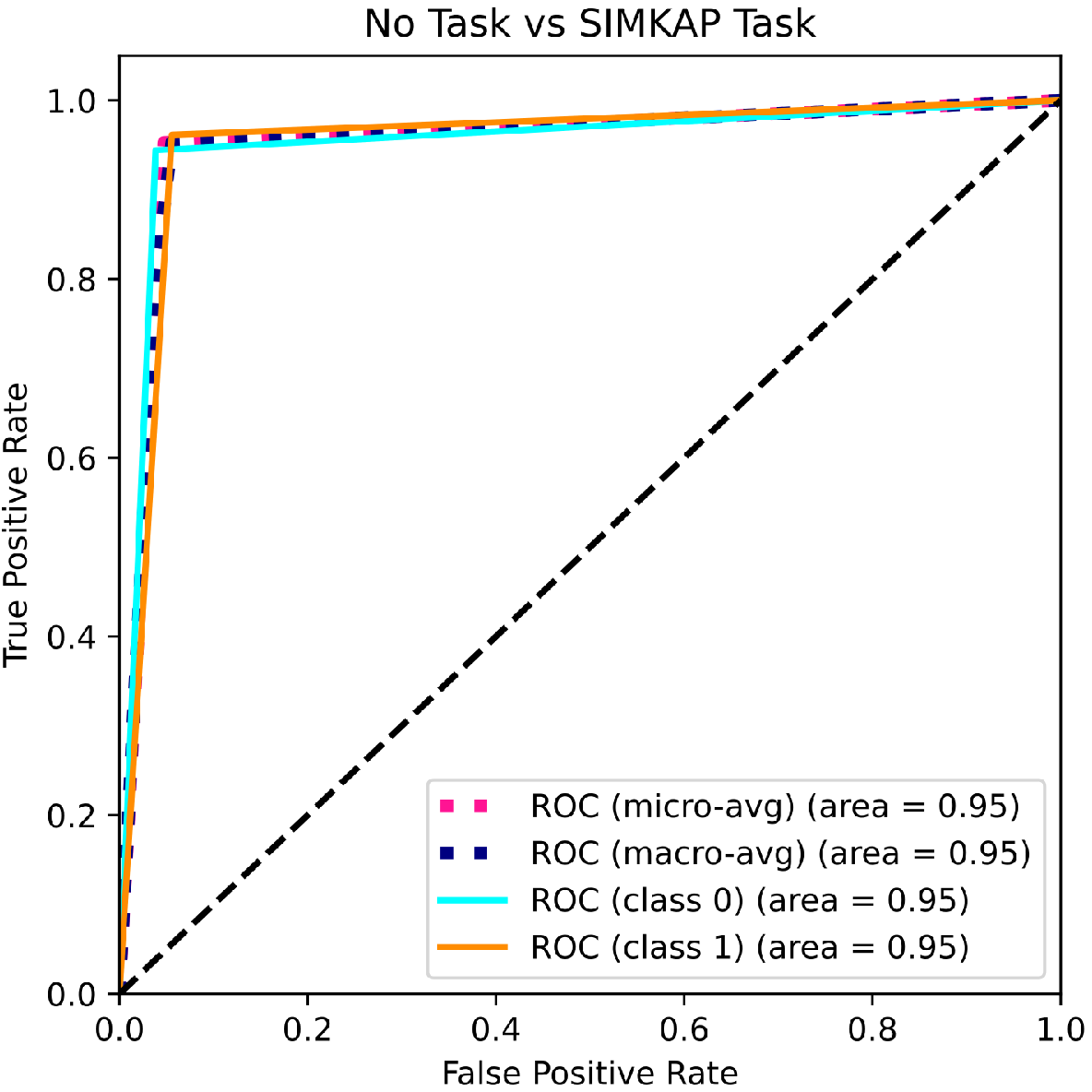}
		\vspace{-0.5cm}
		\caption{}
		\label{fig_roc:is2c}
	\end{subfigure}
	\begin{subfigure}[b]{0.24\textwidth}
	    \centering
		\includegraphics[width=\textwidth]{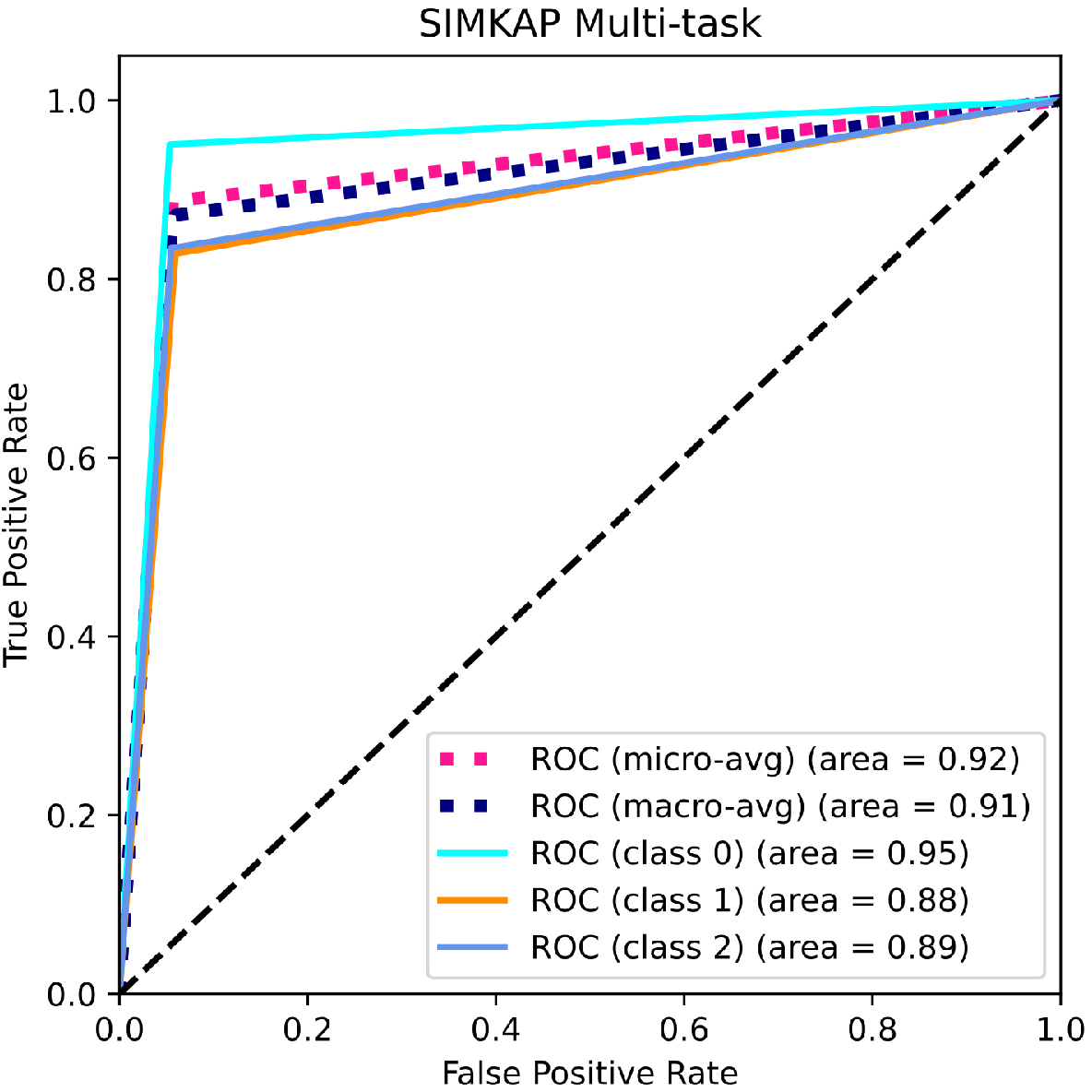}
		\vspace{-0.5cm}
		\caption{}
		\label{fig_roc:is3c}
	\end{subfigure}
	\caption{ROC for Age and Gender dataset (a) Gender, (b) Age; and STEW dataset (c) No vs SIMKAP task, (d) SIMKAP multi-task.}
	\label{fig_roc}
\end{figure*}

\begin{figure*}[!t]
	\centering
	\begin{subfigure}[b]{0.48\textwidth}
	    \centering
		\includegraphics[width=\textwidth]{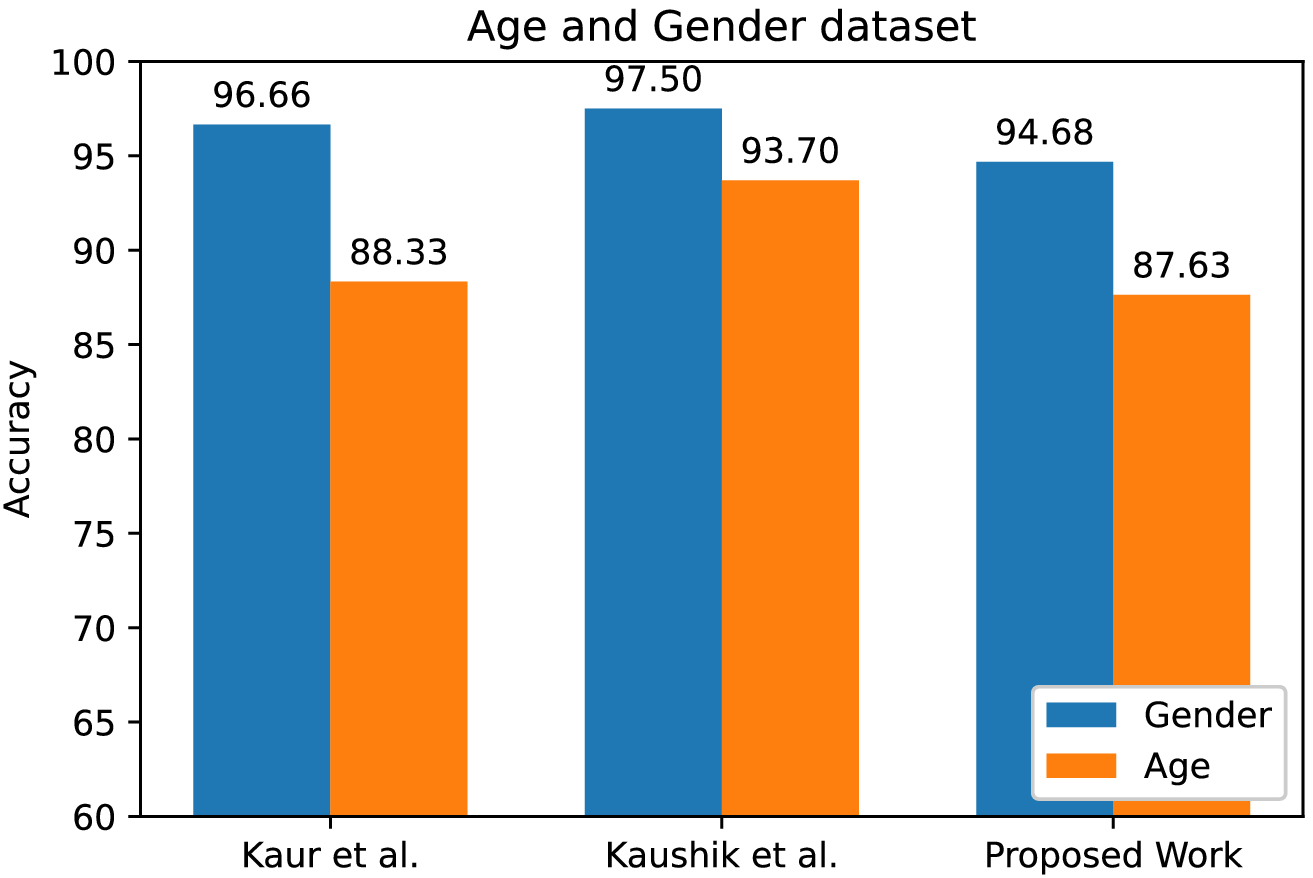}
		\vspace{-0.5cm}
		\caption{}
		\label{fig_bc:bc_ag}
	\end{subfigure}
	\begin{subfigure}[b]{0.48\textwidth}
	    \centering
		\includegraphics[width=\textwidth]{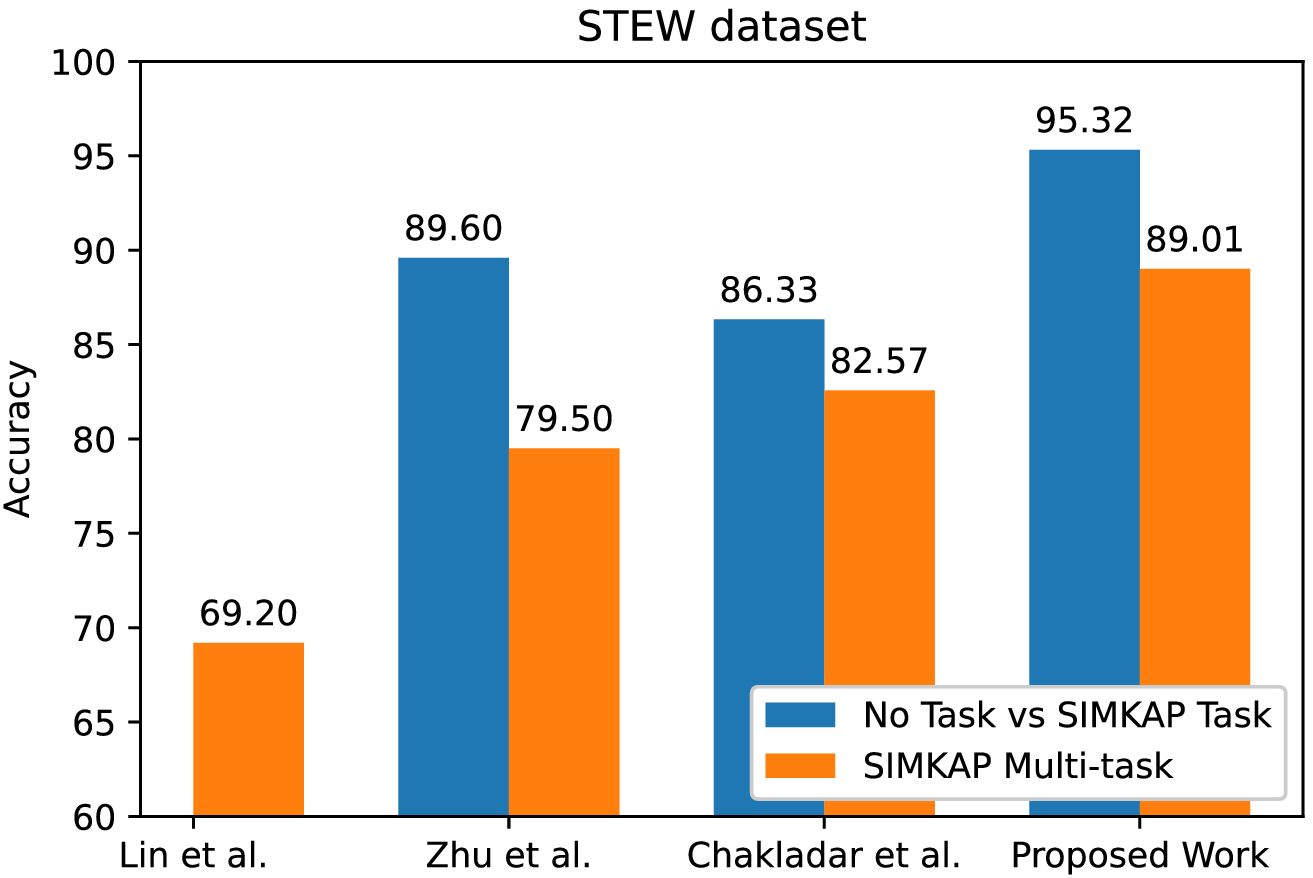}
		\vspace{-0.5cm}
		\caption{}
		\label{fig_bc:bc_is}
	\end{subfigure}
	\vfill
	\vspace*{0.25cm}
	\begin{subfigure}[b]{0.48\textwidth}
	    \centering
		\includegraphics[width=\textwidth]{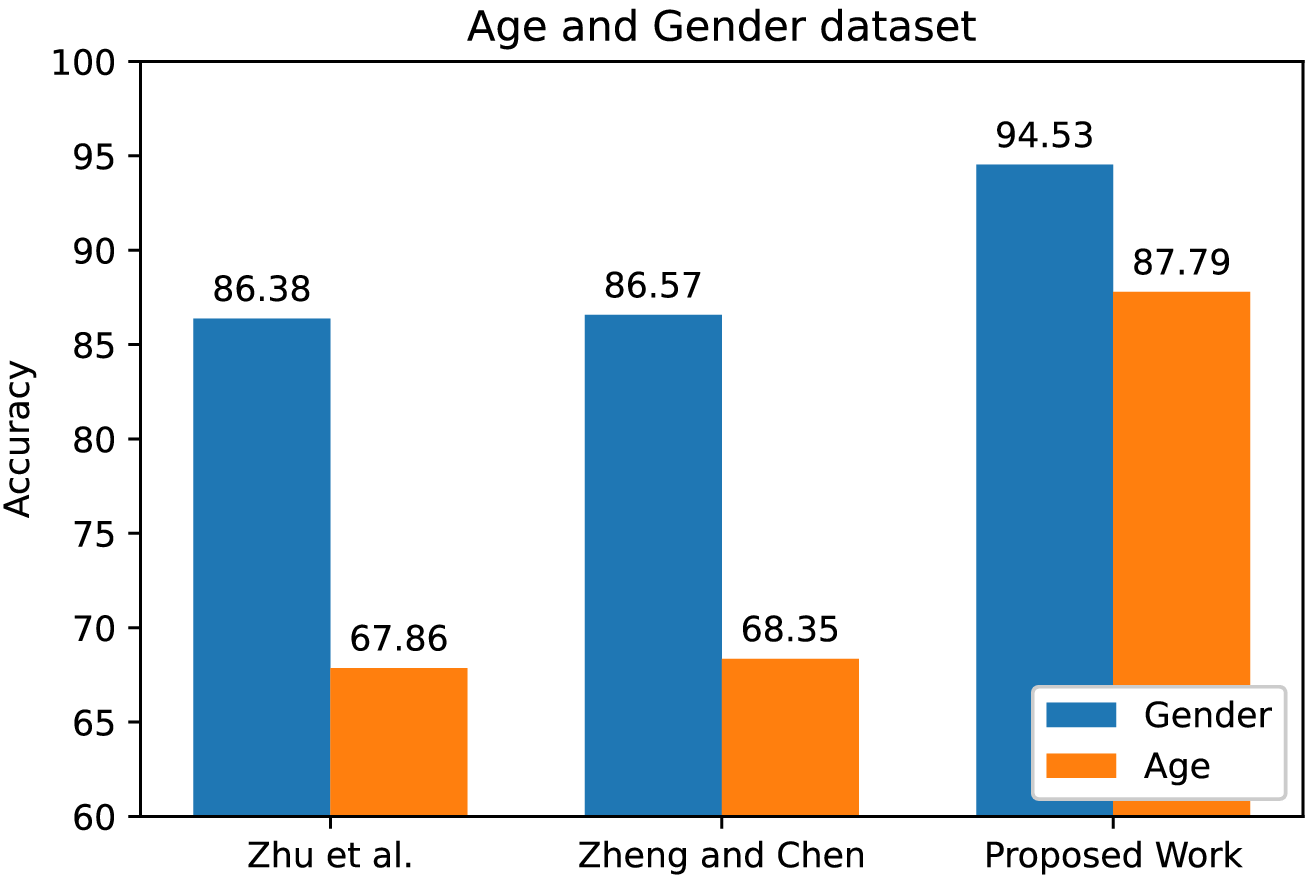}
		\vspace{-0.5cm}
		\caption{}
		\label{fig_bc:abl_ag}
	\end{subfigure}
	\begin{subfigure}[b]{0.48\textwidth}
	    \centering
		\includegraphics[width=\textwidth]{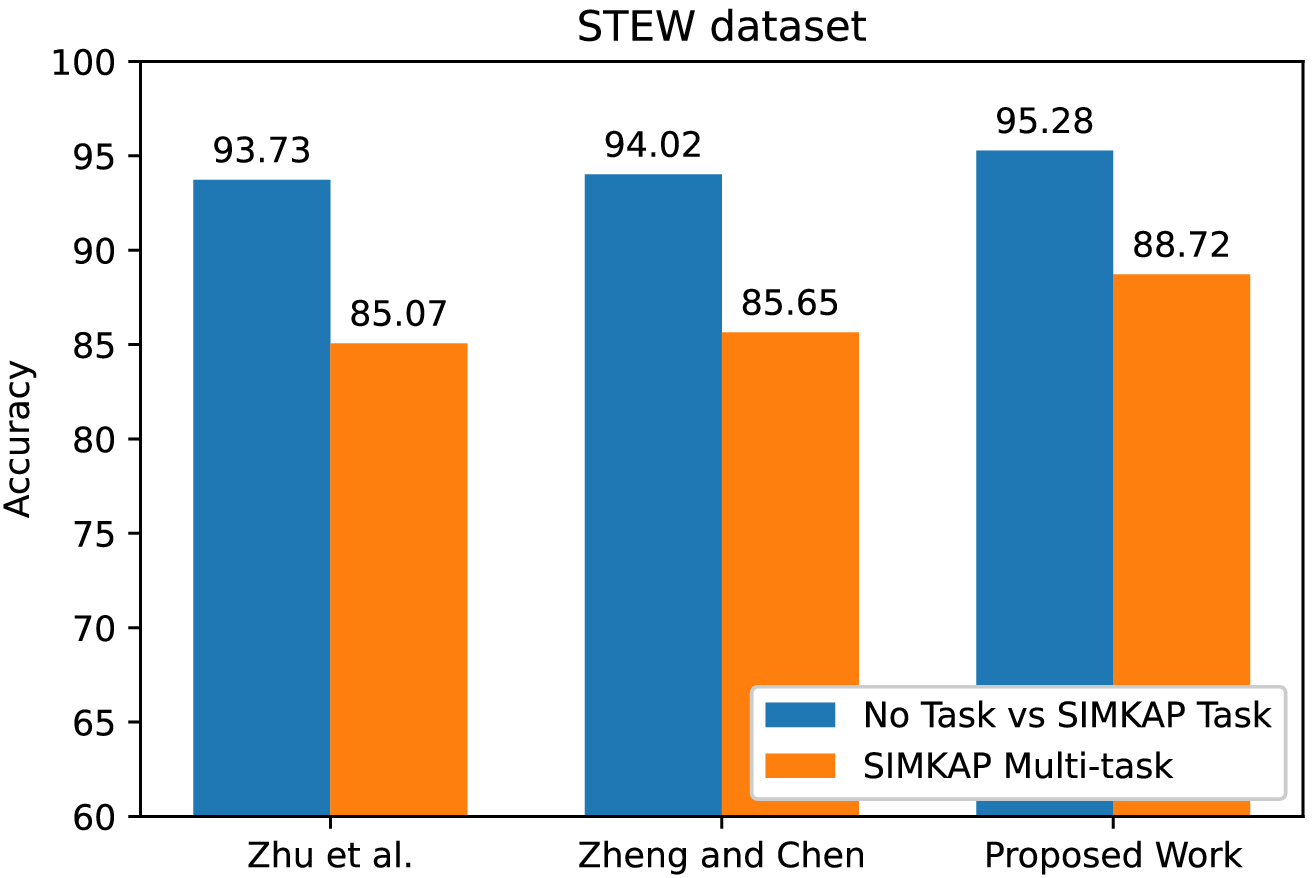}
		\vspace{-0.5cm}
		\caption{}
		\label{fig_bc:abl_is}
	\end{subfigure}
	\caption{Comparison of methods for (a) Age and Gender and (b) STEW dataset; and Ablation Study for (c) Age and Gender and (d) STEW dataset.}
	\label{fig_bc}
\end{figure*}

\begin{table*}[!t]
	\centering
	\caption{Ablation Study}
	\renewcommand\arraystretch{1.1}
	\begin{tabular}{l c c c c c c c c c}
		\toprule
		\textbf{Model} & \textbf{Method} & \multicolumn{2}{c}{\textbf{Accuracy}} & \multicolumn{2}{c}{\textbf{Precision}} & \multicolumn{2}{c}{\textbf{Recall}} & \multicolumn{2}{c}{\textbf{F1-score}} \\

        \midrule
        Age and Gender & & 2 class & 6 class & 2 class & 6 class & 2 class & 6 class & 2 class & 6 class \\
		\midrule
		
		Zhu et al. \cite{zhu2020convolution}
		& CNN
		& 86.38 & 67.86
		& 87.72 & 68.88
		& 80.21 & 68.41
		& 82.64 & 67.72 \\
		
        Zheng and Chen \cite{zheng2021attention} 
        & Bi-LSTM-AttGW
        & 86.57 & 68.35
        & 85.74 & 68.46
        & 82.25 & 68.25
        & 83.66 & 68.24 \\

        \textbf{Proposed Work}
        & \textbf{Transformer}
        & \textbf{94.53} & \textbf{87.79}
        & \textbf{94.39} & \textbf{88.19}
        & \textbf{92.82} & \textbf{87.82}
        & \textbf{93.55} & \textbf{87.99} \\
        
        \midrule
        \midrule
        STEW & & 2 class & 3 class & 2 class & 3 class & 2 class & 3 class & 2 class & 3 class \\
		\midrule
        
        Zhu et al. \cite{zhu2020convolution} 
        & CNN
        & 93.73 & 85.07
        & 93.74 & 83.51
        & 93.73 & 82.60
        & 93.73 & 82.99 \\
        
        Zheng and Chen \cite{zheng2021attention} 
        & Bi-LSTM-AttGW
        & 94.02 & 85.65
        & 94.02 & 83.94
        & 94.01 & 83.66
        & 94.02 & 83.78 \\
        
        \textbf{Proposed Work}
        & \textbf{Transformer}
        & \textbf{95.28} & \textbf{88.72}
        & \textbf{95.29} & \textbf{87.31}
        & \textbf{95.28} & \textbf{87.12}
        & \textbf{95.28} & \textbf{87.21} \\
        
		\bottomrule
	\end{tabular}
	\label{tab_ablation}
\end{table*}

Embedding and positional encoding are done before sending the input to the encoders. Encoders operate with vectors and therefore, the input is converted into embedding vectors using a fixed representation. Unlike recurrent neural networks (RNN) and LSTM, transformer networks do not have a way to capture relative positions of the input. To provide this contextual information, positional encoding is used with each input vector. Positional encoding is not part of the architecture of the model but the pre-processing. The positional encoding vector is generated to be the same size as the embedding for each input. The positional encoding vector is added to the embedding vector after calculation. The `injected' pattern into the embedding vector allows the algorithm to learn this spatial information. This mechanism is shown in Fig \ref{fig_epe}.


\subsection{Classification}

For Age and Gender classification, the architecture for gender (2 class) classification is shown in Fig \ref{fig_arch:ag}. The last layer is modified as \textit{Dense(6)} for age (6 class) classification network. The number of attention heads is also increased to 8. The rest of the parameters are kept the same. The dataset is evaluated for two classification scenarios. First, for gender classification, the dataset has been categorized based on the gender of the subject, male and female. This network (2 class) is trained for 49 epochs with early stopping (Fig \ref{fig_th:ag2c}). Second, for age classification, the dataset has been categorized into six age groups, 6-11, 12-16, 18-24, 25-30, 33-39, and 42-56. These age groups are then labeled as 0-5. This network (6 class) is trained for 122 epochs with early stopping (Fig \ref{fig_th:ag6c}).

For classification of the STEW dataset, the network architecture for ``No vs SIMKAP task'' is shown in Fig \ref{fig_arch:is}. The last layer is modified as \textit{Dense(3)} for the ``SIMKAP multi-task'' network. For embedding, 32 dimensions are used with 4 attention heads. The hidden layer size of the feed-forward network in the transformer is 64. Adam is used as an optimizer with default parameters and sparse categorical cross-entropy is used for loss. Both datasets are used as a 70-15-15 percent split for train, validation, and test sets during the experimentation. STEW dataset is evaluated for two classification scenarios. First, for ``No vs SIMKAP task'', the dataset has been categorized into two workload levels, at rest as low and performing SIMKAP tasks as high. This network (2 class) is trained for 51 epochs with early stopping (Fig \ref{fig_th:is2c}). Second, for the ``SIMKAP multi-task'', the dataset is categorized into three workload levels, using the ratings provided by each subject. One can perceive a rating of 1-3 as low, 4-6 as moderate, and 7-9 as high workload. This network (3 class) is trained for 73 epochs with early stopping (Fig \ref{fig_th:is3c}).


\section{Results and Discussion}
\label{sec:results}

This section discusses the comparative analysis between the proposed model and other approaches, including machine learning and popular deep neural networks. The results are summarized in Table \ref{tab_results}, confusion matrices are shown in Fig \ref{fig_cm}, receiver operating characteristics (ROC) are shown in Fig \ref{fig_roc}, comparative analysis is shown in Fig \ref{fig_bc}, and ablation study is summarized in Table \ref{tab_ablation}. Also, the features generated from the network corresponding to their input embedding can be seen in Fig \ref{fig_va} for both classes of both the datasets.



\subsection{Age and Gender Dataset}%

The proposed work make use of transformer network (Fig \ref{fig_arch:ag}) on raw EEG data and achieved an accuracy of 94.68\% and 87.63\% for gender and age classification, respectively (Fig \ref{fig_bc:bc_ag}). The corresponding confusion matrices are shown in Fig \ref{fig_cm:ag2c} and \ref{fig_cm:ag6c} respectively and ROC are shown in Fig \ref{fig_roc:ag2c} and \ref{fig_roc:ag6c} respectively. The accuracy achieved with the transformer network trained on pre-processed data is not far behind the state-of-the-art accuracy. The reason could be the positional encoding, which can be done in different ways. The positional encoding used in transformer networks for text processing is not tailor-made for EEG data. Also, the features are not used, which are more effective in some cases. (Fig \ref{fig_va:ag2c} and \ref{fig_va:ag6c}).


\subsection{STEW Dataset}

The proposed work make use of transformer network (Fig \ref{fig_arch:is}) for classification and achieved test accuracy of 95.32\% and 89.01\% for ``No vs SIMKAP task'' and ``SIMKAP multi-task'' classification on raw EEG data. The corresponding confusion matrices are shown in Fig \ref{fig_cm:is2c} and \ref{fig_cm:is3c} respectively and ROC are shown in Fig \ref{fig_roc:is2c} and \ref{fig_roc:is3c} respectively. Transformer network trained with pre-processed EEG data clearly out-performs the existing methods with feature extraction for both classification scenarios on the STEW dataset (Fig \ref{fig_bc:bc_is}). One of the reasons could be the effective learning of embedding and positional encoding used which provides contextual information.


\subsection{Ablation Study}

For the ablation study, highly successful networks proposed by Zhu et al. \cite{zhu2020convolution} and Zheng and Chen \cite{zheng2021attention} are used. These are trained on both datasets used in this work. The accuracies achieved by the proposed model achieved better results than both the methods for both datasets. The improved accuracy might result from multi-head attention, which learns better and faster than a single attention layer used in these networks, resulting in high performance. The comparison of results with these attention-based networks is summarised in Table \ref{tab_ablation}. The comparison is also shown in Fig \ref{fig_bc:abl_ag} and \ref{fig_bc:abl_is} for Age and Gender and STEW dataset respectively. It can be seen that the proposed transformer network for EEG outperforms existing deep learning networks on raw EEG data.


\section{Conclusion}
\label{sec:conclusion}

This study explored the use of transformer networks for the classification of raw (cleaned) EEG data. To evaluate such models' efficacy, proposed network's performance was tested on a local (Age and Gender) and a public dataset (STEW). The robustness of the classifier was also tested as both datasets had different classification paradigms. Accuracies of 94.68\% (gender), 87.63\% (age), 95.59\% (two workload levels), and 89.40\% (three workload levels) were achieved. The proposed method gave state-of-the-art accuracy for STEW and comparable for Age-and-gender. This proves the effectiveness of the transformers for processing raw EEG data and can reduce the need for feature extraction in EEG as deep learning has done in other domains such as image processing. 

However, the study needs to be further replicated and validated with different datasets and more comparisons. The positional encoding used in transformers is not tailor-made for EEG data and features are not used, which are more effective in some cases. Both of these may be the reason for lower performance on the Age and Gender dataset. It can be further improved by generating robust EEG embedding using multiple datasets similar to the embedding used in NLP. Similarly, new positional encoding methods specifically to EEG can be explored in the future. In conclusion, transformer networks seem to be effective in classifying EEG data as well. They might also reduce the over-reliance on the feature extraction still used in the EEG domain.


\normalsize
\bibliography{references}


\end{document}